\documentclass[10pt,pre,notitlepage,twocolumn,superscriptaddress,color,showpacs]{revtex4-1}
\usepackage{url,hyperref,lineno,microtype,epsf}
\usepackage{amsmath,amsfonts,amssymb}
\usepackage{graphicx}
\usepackage[T1]{fontenc}

\begin{document}
\title{Coevolutionary Dynamics of Group Interactions: Coevolving Nonlinear Voter Models}
\author{Byungjoon Min}
\affiliation{Department of Physics, Chungbuk National University, Cheongju, Chungbuk 28644, Korea}
\email{bmin@cbnu.ac.kr}

\begin{abstract}
We survey the coevolutionary dynamics of network topology and group interactions in opinion 
formation, grounded on a coevolving nonlinear voter model. The coevolving nonlinear voter 
model incorporates two mechanisms: group interactions implemented through nonlinearity in 
the voter model and network plasticity demonstrated as the rewiring of links to remove 
connections between nodes in different opinions. We show that the role of group 
interactions, implemented by the nonlinearity can significantly impact both the dynamical 
outcomes of nodes' state and the network topology. Additionally, we review several 
variants of the coevolving nonlinear voter model considering different rewiring
mechanisms, noise of flipping nodes' state, and multilayer structures. We portray the various 
aspects of the coevolving nonlinear voter model as an example of network coevolution driven 
by group interactions, and finally, present the implications and potential directions for 
future research.
\end{abstract}

\maketitle

\section{Introduction}

The structures of complex networks that govern interactions between individuals have an essential impact 
on the dynamics of those agents~\cite{book_newman,boccaletti}. 
The network structures not only affect the dynamical processes taking place within the system, but 
they themselves also evolve under the influence of individual states~\cite{evolution,temporal}.
In this regards, the integration of the dynamics of individual states and network structures
has received much attention from physics, social science, and network science communities 
~\cite{zimmermann2001,zimmermann2004,holme2006,gross2006,vazquez2008,gross2008,adaptive_book}.
Researchers have explored the coevolutionary dynamics of complex networks in various contexts
by integrating the evolution of networks' topology and the dynamics of nodes~\cite{adaptive_book}. 
There have been many studies on coevolving or adaptive networks, such as coevolving voter models 
~\cite{holme2006,vazquez2008,nardini2008,durrett2012,carro2014,diakonova2014,diakonova2015,saeedian2019}, 
coevolving spin systems~\cite{biely2009,mandra2009,raducha_spin,korbel2023}, coevolutionary 
opinion dynamics~\cite{kimura2008,sudo2013,liu2023}, language evolution and competition 
~\cite{carro2016}, adaptive epidemic models~\cite{gross2006,shaw2008,demirel2014,scarpino2016,saeedian2017,
marceau2018,achterberg2020,choi2023}, cultural evolution models~\cite{coevAxelrod1,coevAxelrod2},
game theoretical models~\cite{ebel2002,eguiluz2005,perc2013,gonzalez2023}, and
biological evolutions~\cite{drossel2001,liu2006,fialkowski2023}.
Owing to the feedback inherent in adaptive systems where the network structure 
influences the node dynamics and vice versa, these studies have revealed intriguing properties 
such as the dynamical organization of network structures and diverse phase 
transitions~\cite{vazquez2008,sudo2013,liu2023,coevAxelrod1}.

The majority of existing research on coevolving networks has largely focused on systems that 
change dynamically according to pair-wise interactions 
~\cite{holme2006,gross2006,vazquez2008,gross2008,adaptive_book}.
However, in various social, neural, biological, and ecological systems, however, group or collective 
interactions are commonly observed~\cite{schelling,granovetter,centola2007,centola2010,
levine2017,monsted2017,bmin_complex,bmin2018_dual,centola2018,laurent2020,battiston2021}
Group interaction refers to a process where more than two nodes participate 
simultaneously, rather than interactions occurring solely between pairs~\cite{battiston2020}. 
Such group interactions can either be composed of collections of pair-wise interactions or 
have the form of many-body interactions that cannot be reduced to mere pair-wise 
interactions~\cite{battiston2021,battiston2020}.
Understanding the group interactions is vital to predict and to control the behavior 
of complex systems~\cite{watts2002,laurent2020,battiston2021}.

While these group interactions are common in reality, there is a gap 
addressing the coevolutionary dynamics driven by group interactions, pointing to the need for further 
examination in this area. Despite the initial research on the coevolutionary dynamics on networks based 
on group interactions for evolutionary games~\cite{perc2013,alvarez2021}, spin models~\cite{mandra2009},
contagion dynamics~\cite{lambiotte2011,bmin2023}, and voter models~\cite{bmin2018}, there remain
several issues for more extensive and thorough studies to grasp its full implications.
An interesting example of coevolutionary dynamics with group interactions would be a coevolving 
nonlinear voter model (CNVM)~\cite{bmin2018,kereh2020}. This model is an extension of a voter model 
which is a well established model based on pair-wise interactions~\cite{holme2006,vazquez2008}, 
to include group interactions~\cite{castellano2009,schweitzer2009,jedrzejewski2017,peralta2018}.

In this review article, we explore the dynamical consequences of the CNVM as a case study
of combining group interactions and coevolutionary dynamics.
We begin with the background of the CNVM, review its fundamental results, and also explore 
several variants of the model~\cite{raducha,bmin2019,raducha2020,jedrzejewski2020}.
This paper aims to provide an example for analyzing the effect of group interactions 
in coevolving networks and deepens our understanding within the broader landscape of 
coevolutionary dynamics.

The paper is organized as follows. First, as a benchmark for comparison, we introduce the 
coevolving voter model with pair-wise interactions in Sec. II. Next, we present 
the coevolving ``nonlinear'' voter model encoded with group interactions, outlining its 
fundamental results in Sec. III. We then sequentially introduce several variants of 
this model with incorporating triadic closure, rewire-to-random mechanism, noise, and 
multilayer coevolution in Sec.~IV. Finally, summary and outlook are presented in Sec. V.

\section{Coevolutionary dynamics of the voter model}

The coevolving voter model is a compelling framework for understanding 
the coevolution of node states and network topology.
This simple yet insightful model describes the evolution of node states under voter 
dynamics and the concurrent reorganization of the network through link rewiring~\cite{vazquez2008}. 
That is, unlike the classical voter model that assumes static connections 
among individuals, the coevolving voter model takes into account the change of 
social ties under the influence of the state of nodes.

The coevolving voter model consists of two dynamical processes: copying and 
rewiring~\cite{holme2006,vazquez2008,durrett2012}.
A node can adopt the opinion of one of its neighbors chosen randomly.
Concurrently, the connections between voters can be rewired, depending on the agreement 
or disagreement of states between two connected voters. 
The specific rule of the coevolving voter model is as follows:
Initially, each node can be one of two states, called up or down, on a network.
In this paper, we denote the state of node $i$ as $\sigma_i$, and thus 
$\sigma_i \in \{-1,1 \}$ where $-1$ and $1$, respectively corresponds to
down and up state. At each update step, we choose one node, say $i$, at random. 
Then with a complementary probability $1-p$, node $i$ copies the state of one 
of its neighbors, for instance $j$, that is selected at random. 
With a probability $p$, if the states of node $i$ and $j$ are different,
node $i$ cuts its connection to node $j$ and establishes a new connection to a node that 
has the ``same'' state with node $i$. 
The procedures continue until the system reaches a steady state.

The parameter $p$, called as network plasticity, represents the ratio between 
the link rewiring and copying processes, thereby determining the time scale between them. 
Depending on the value of $p$, there are two distinct phases, an active to a frozen phases,
in the steady state. At a critical point $p_c$, the generic absorbing phase transition occurs 
from the active to the frozen phase~\cite{vazquez2008}. 
The active phase corresponds to a connected network at a steady state 
in the thermodynamic limit, and is characterized by a non-zero density of active links. 
On the other hand, the frozen phase corresponds to a fragmented network where 
each component is in a consensus state. 
The fragmentation transition between active and fragmented phases is a 
peculiar characteristic of the coevolving voter model.

\section{Coevolving Nonlinear Voter Model}

The coevolving voter model provides an important insight into many interesting 
phenomena, such as polarization or consensus in society~\cite{vazquez2008} and language 
competition~\cite{carro2016}. In addition, there have been many variants of the coevolving 
voter model for more realistic modeling, i.e., incorporating noise in the flipping of 
opinions~\cite{diakonova2015}, multilayer coevolution~\cite{diakonova2014,klimek2016}, and 
signed interactions~\cite{saeedian2019}.
Along the studies on the coevolving voter models, it is steadily assumed that the dynamical
processes are governed by dyadic interactions where the state of a node or network topology
change based on the influence of pairs~\cite{carro2014,diakonova2014,diakonova2015,saeedian2019}. 
However, this simple approach fails to capture the complexity of real-world social 
dynamics where multiple agents collectively influence an individual~\cite{schelling,granovetter,
centola2010,levine2017,monsted2017,battiston2021}. 
This suggests that an agent engages in ``nonlinear'' interactions with its neighbors, 
to implement group interactions in the voter model~\cite{castellano2009,schweitzer2009,peralta2018}.

\begin{figure}
\includegraphics[width=\linewidth]{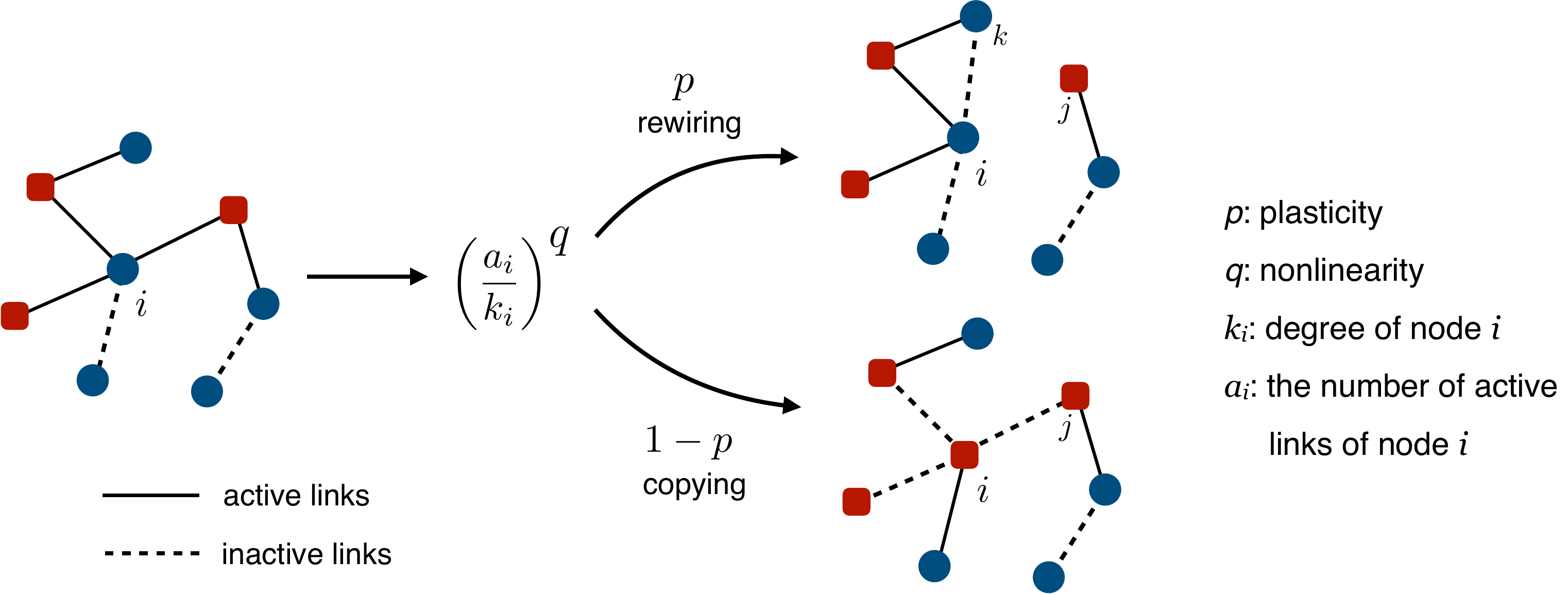}
\label{fig1}
\caption{ 
Schematic illustration of dynamical rules of the coevolving nonlinear voter model.
Nodes are either up (red circle) or down (blue square). 
Dashed lines are active links while solid ones represent inactive links. 
At each step, we randomly select node $i$, then a neighbor $j$ connected through 
an active link with probability $\left( \frac{a_i}{k_i} \right)^q$, where 
$a_i$ is the number of active links. 
With probability $p$, node $i$ removes its connection with $j$
and forms a new link with a node that has the same opinion as $i$. 
On the other hand, with probability $1-p$ node $i$
changes its state to imitate that of node $j$.
}
\end{figure}

To address the limitations of pairwise interactions in coevolving voter models, a coevolving nonlinear 
voter model that combines evolutionary dynamics of networks with nonlinear interactions between agents
was proposed~\cite{bmin2018}. The CNVM is an extension on coevolving networks of a nonlinear voter 
model~\cite{castellano2009,nyczka2012,mobilia2015,peralta2018}.
The dynamical rules of the CNVM are as follows (Fig.~1). 
At each time step, we choose a node, say $i$, at random. 
Next, we measure the fraction $\pi_i$ of active links of node $i$ to its degree as
\begin{linenomath} 
\begin{align}
\pi_i = \frac{a_i}{k_i},
\end{align}	
\end{linenomath} 
where $a_i$ represents the number of active links of node $i$. Here, active links
means links that connect a pair of two nodes with different opinions. 
We then choose at random an active link to a neighbor, say $j$ with a probability, 
$\pi_i^q$, where $q$ is the degree of nonlinearity. 
And, with the complementary probability $1-\pi_i^q$,
nothing happens. Upon selecting nodes $i$ and $j$, we proceed coevolutionary dynamics 
with network plasticity $p$. Specifically, with a probability $p$, node $i$ disconnects its 
link to $j$ and establishes a new link with another node
that shares the same opinion as node $i$. Conversely, with 
a probability of $1-p$, node $i$ changes its state to imitate the state of node $j$. 
This process continues until the system reaches an active steady state or an absorbing state.
Note that the network structure changes over time while maintaining 
a constant link density of network, that is a constant average degree $\langle k \rangle$.

In the CNVM, there are two important parameters: network plasticity $p$ and nonlinearity $q$.
The network plasticity $p$ represents the rate of link rewiring.
The degree of nonlinearity $q$ is a parameter newly introduced in the CNVM
to represent group interactions. To understand the role of this parameter, consider 
the following scenario. Suppose a voter, rather than asking a single neighbor, inquires about
the opinions of multiple neighbors, specifically $q$ neighbors. And, if all these $q$ neighbors 
have the same opinion, then the voter does change its opinion. To be more precise, our model 
corresponds to the scenario when a voter randomly chooses one out of the neighbors $q$ times,
allowing repetition. If we mathematically describe it, we arrive at the introduced term 
$\pi_i^q$. Unlike the ordinary voter model, it 
reflects group interactions because it requires convergence of the opinions of multiple 
neighbors.

Let us examine the effect of $q$ qualitatively. For the linear case, when 
$q=1$, it corresponds to selecting an active link randomly, which is exactly the same
as the ordinary coevolving voter model. If $q>1$, the ratio between the probability of following 
the majority opinion and minority opinion among neighbors is higher compared to the linear voter model.
Conversely, when $q<1$, the probability of following with the minority opinion becomes 
relatively higher.
The empirical evidences of the nonlinear interactions can be found in 
social impact theory~\cite{nowak}, language competition, and extinction~\cite{nettle,abrams}.
While it is observed that $q<1$ in social impact theory~\cite{nowak} and language 
evolution~\cite{nettle}, it was found to $q=1.3$ in language extinction processes~\cite{abrams}.

Introducing the nonlinearity $q$ as an additional dimension, the CNVM exhibits a rich variety of 
phenomena. As shown in Fig.~2(a), depending on the values of network plasticity $p$ and 
nonlinearity $q$, there are three possible phases that can reach in the steady state: 
consensus, fragmentation, and coexistence.
In the consensus phase, all nodes arrive at the same state, either up or down.
This is an absorbing state, meaning that once reached the system remains in this phase permanently. 
The fragmentation phase, on the other hand, represents a situation where a network breaks 
into multiple components of localized consensus. In this phase, agents within a particular component
share the same opinion, but this opinion may differ from component to component. 
Finally, the coexistence phase is characterized by the persistent presence of both opinions
throughout the system without the formation of disconnected components.
This phase is a dynamically active state where two different opinions exist in the same component
and continuously interact each other.

\begin{figure}[t!]
\includegraphics[width=\linewidth]{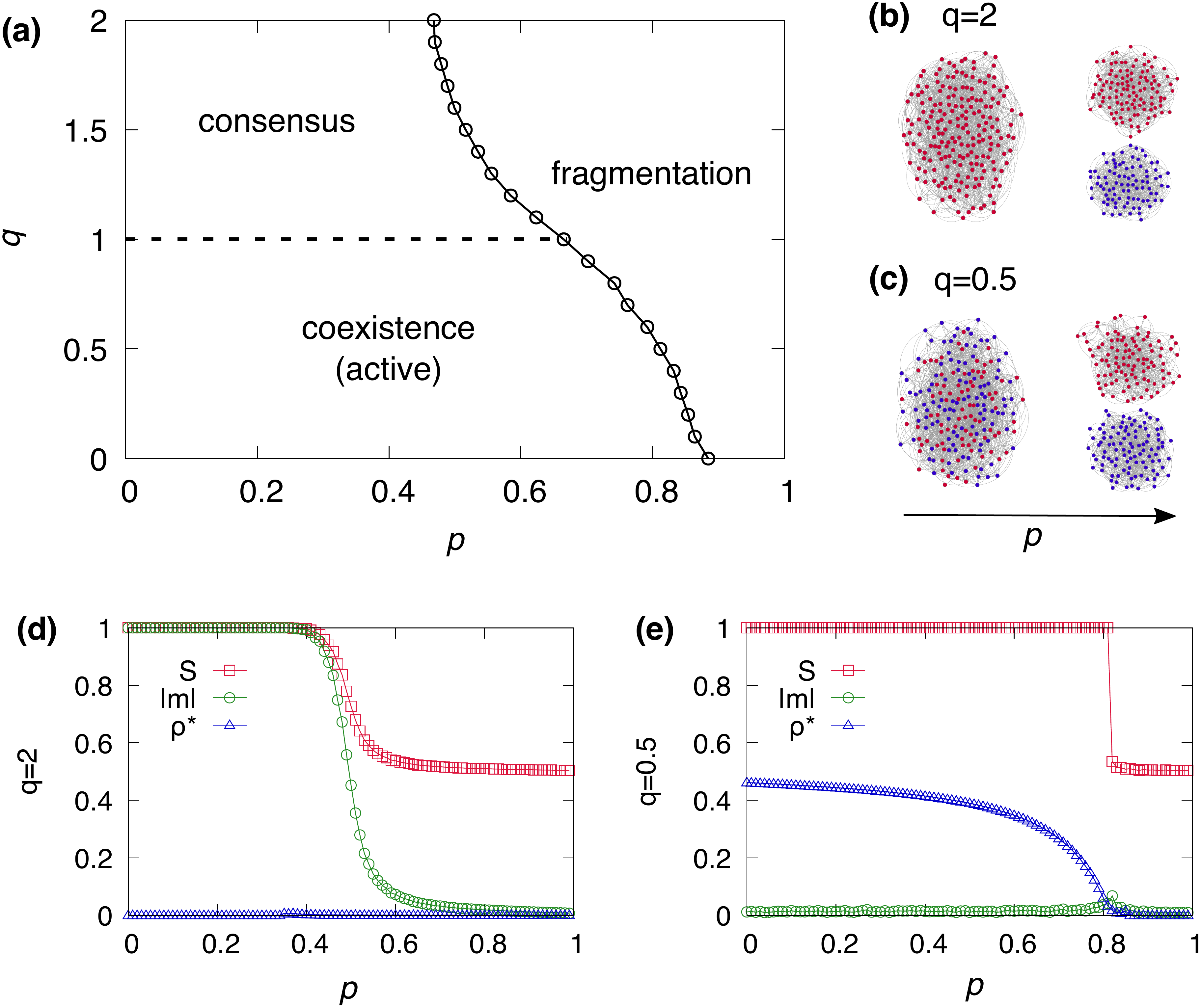}
\label{fig2}
\caption{
(a) Phase diagram of the CNVM with respect to $p$ and $q$ contains consensus, 
coexistence, and fragmentation phases. The diagram was obtained numerically from 
degree regular networks with $\langle k\rangle=8$, $N=10^4$, and initially neural 
magnetization $m=0$.
(b,c) Schematic illustrations of phase transitions:
(b) transitions from consensus to fragmentation phases for $q=2$
and (c) transitions from coexistence to fragmentation phases for $q=0.5$.
(d,e) Numerical results of the size of the largest component $S$, the absolute 
value of magnetization $|m|$, and the density of active links $\rho$ as a function 
of $p$ are shown together: (d) $q=2$ and (e) $q=0.5$, on degree regular networks 
with $\langle k\rangle=8$, $N=10^4$, and initially neural magnetization $m=0$.
}
\end{figure}

The three phases can be characterized by the quantities: the size of the largest component $S$, 
the absolute value of magnetization $|m|$, and the density of active links $\rho$. 
The magnetization $m$ refers to the average value of the state of nodes, defined by
\begin{linenomath} 
\begin{align}
m=\frac{1}{N} \sum_i \sigma_i,
\end{align}
\end{linenomath} 
where $N$ is the total number of nodes and $\sigma_i \in \{-1,1 \}$. 
The consensus phase corresponds to $(S,|m|,\rho) \approx (1,1,0)$ 
since there is a single component with the same state of nodes.
The fragmentation phases corresponds to $(1/2,0,0)$, meaning that 
there are two separated components where agents within a component share the same opinion.
Finally, the dynamically active coexistence phase is characterized by $(1,0,\rho^*)$
where $\rho^*$ is a finite value of density of active links.
In the coexistence phase, nodes with opposite opinions exist in the  same component.
Here we assume that the dynamics starts from neutral magnetization, $m=0$.

The CNVM shows a fragmentation transition between connected and disconnected networks 
but with different mechanisms depending on the nonlinearity $q$. 
The transitions to fragmentation with increasing $p$ are qualitatively illustrated in Fig.~2(b), 
which shows the transitions from consensus to fragmentation phases for $q=2$, and in Fig.~2(c), 
which shows the transition from coexistence to fragmentation phases for $q=0.5$.
Figure 2 shows the numerical results for the characteristics of phase transitions 
for (a) $q=2$ and (b) $q=0.5$. The size of the largest component $S$, absolute value 
of magnetization $|m|$, and the density of active links $\rho^*$ at the steady state 
are shown together. 
For $q>1$ there is a transition between consensus and fragmentation phases as shown 
in Fig.~2(d) for $q=2$. Since the transition occurs between two absorbing states, 
it is different from the continuous absorbing transition observed in the coevolving 
linear voter model.
However, for $q<1$ the system undergoes a continuous transition between a dynamically 
active phase and a fragmentation phase [Fig.~2(e)], similar to the linear case.


The behaviors of the CNVM can be understood by the coupled differential equations of the 
magnetization $m$ and density of active links $\rho$. 
Based on the pair approximation, the coupled equations on a random network with 
average degree $\langle k \rangle$ can be derived as~\cite{bmin2018}
\begin{align}
\label{eq:mf1}
\frac{dm}{dt}&=2(1-p)\left[ - n_+ \left( \frac{\rho}{2 n_+} \right)^q + n_{-}\left( \frac{\rho}{2 n_{-})} \right)^q \right],\\
\label{eq:mf2}
\frac{d \rho}{dt}&= \frac{2}{\langle k \rangle} \Bigg\{ 
-p\left[ n_+ \left( \frac{\rho}{2n_+} \right)^q +n_{-}	\left( \frac{\rho}{2 n_{-} }\right)^q \right]  \\
+& (1-p)\Bigg[ n_+ \left(\frac{\rho}{2 n_+}\right)^q \left(\langle k\rangle -2 q -2 (\langle k \rangle -q) \frac{\rho}{2 n_+} \right) \nonumber \\
+&n_{-} \left(\frac{\rho}{2 n_{-}}\right)^q \left(\langle k\rangle -2 q -2 (\langle k \rangle -q) \frac{\rho}{2 n_{-}} \right)\Bigg] \Bigg\}, \nonumber
\end{align}
where $n_+=(1+m)/2$ and $n_-=(1-m)/2$.
In this approximation, a homogeneous structure of networks is assumed. 
The rationale of each term in the equations is following.
If we select node $i$ with state $s$ where $s \in \{+1,-1 \}$, the probability that its connected neighbor has a different 
state can be estimated by $\rho/(2n_s)$, where $n_s$ stands for the fraction of nodes in state $s$
and $s$ can be either $+1$ or $-1$. In addition, if node $i$ in state $s$ is selected 
for an update to be a different state with a probability of $[\rho/(2 n_s)]^q$, 
then it is estimated that $q$ of its neighbors are in a different state. 
Thus, the remaining neighbors, denoted as $\langle k \rangle - q$, will have 
a different state with a probability of $\rho/(2 n_s)$~\cite{bmin2018}.

The coupled equations based on the pair approximation provide a phenomenological 
explanation for the phases and phase transitions observed in the CNVM model~\cite{bmin2018}.
Based on the approximation, the transition point $p_c$ is predicted as 
\begin{linenomath} 
\begin{align}
p_c = \frac{\langle k \rangle - 2q}{1 + \langle k \rangle -2q}.
\end{align}
\end{linenomath} 
For $p>p_c$, the fragmented phase $(m,\rho)=(0,0)$ is a stable solution across all $q$ values. 
On the other hand, for $p<p_c$, the system shows a dynamic active phase or an absorbing phase 
with a connected network structure, depending on the value of $q$.
When $q<1$, the solution $(0,\rho^*)$ becomes stable, suggesting 
a dynamically active phase. Conversely, when $q>1$ the solutions $(-1,0)$ and $(1,0)$ become
stable, indicating an absorbing consensus phase.

\section{Variants of Coevolving Nonlinear Voter Model}

The ``nonlinearity'' in the CNVM model offers the group interactions that
governs the change of state and network structure, beyond pair-wise interactions.
For more realistic approaches, there have been several extensions in the CNVM.
Among the generalized models, we focus on the effect of the triadic closure, 
rewire-to-random mechanism, noise, and multilayer coevolution. Triadic closure 
stands for the mechanism forming new connections with the neighbors 
of one's neighbors, a departure from the standard random rewiring approach.
Next, we discuss the difference between the rewire-to-same and rewire-to-random mechanism
when we find a new neighbor during link rewiring.
Noise in the CNVM means that individuals can randomly flip their opinions, irrespective 
of their neighbors' states, introducing an unpredictable element to the dynamics. 
Finally, multilayer coevolution represents the coevolutionary dynamics 
within networks composed of multiple layers, examining how interactions 
between layers affect the dynamical consequences.
In the following sections, we discuss the variations of the CNVM 
focusing on the difference to the ordinary CNVM.

\subsection{Coevolving Nonlinear Voter Model with Triadic Closures}

It is common in social systems to form ties locally when seeking new 
connections~\cite{park2003,lee2010,klimek2013}. Triadic closure is aptly captured 
by the local evolution of network structures. In other words, when individuals are on the 
look for new connections, they tend to connect with the neighbors of their current neighbors,
to form a triangle. A natural and straightforward extension of the CNVM is 
to implement this triadic closure mechanism in link rewiring~\cite{raducha}.
Specifically, a node attempts to find a new neighbor among the neighbors of their 
current neighbors during link rewiring~\cite{raducha,malik2016}. This rewiring 
pattern reflects many real-world networks, especially social networks where 
acquaintances of acquaintances often become directly connected~\cite{park2003,lee2010}.

A peculiar phenomenon with triadic closures compared to the CNVM 
is a shattered phase, which appears when $q<1$ and $p<p_c$~\cite{raducha}. 
In this phase, the system remains an active phase with zero magnetization, 
$|m|=0$, like the coexistence phase in the CNVM. However, the structure of 
networks consists of a large active component alongside numerous isolated 
nodes, so called shattered phase. It implies for $q<1$ with triadic closure 
that as $p$ decreases a network, initially characterized by two separated 
components with opposite opinions, evolves into many isolated nodes and an 
active component, so called a shattering transition. In addition, the clustering 
coefficient shows values that are not close to zero, while it consistently 
approaches zero for all parameter sets for random link rewiring in the ordinary CNVM.

\subsection{Coevolving Nonlinear Voter Model with Rewire-to-Random Mechanism}

In the CNVM, when searching for new neighbors through rewiring, a new link  is
established with a neighbor in the same state. However, in real-world social 
systems, one may not precisely identify the opinions 
or states of other agents. Therefore, it may be natural to rewire with any random 
node from the entire network, regardless of its state~\cite{durrett2012,jedrzejewski2020}. 
The specific model is as follows: A node $i$ is randomly 
selected, and its state or link is changed depending on the nonlinearity $q$, similar 
to the previous model. However, when rewiring, a new neighbor is randomly chosen from 
the entire network, irrespective of its current state. Therefore with the modification
a newly connected neighbor can have a different opinion.

The effect of the rewire-to-random mechanism was examined using the pair approximation
and numerical simulations~\cite{jedrzejewski2020}. 
The modified model produces two dynamically active phases, symmetric and asymmetric, 
in addition to consensus and fragmentation. The active symmetric 
phase exhibits the same number of nodes in two states, indicating no preferred 
state in the network. Conversely, the active asymmetric phase shows a dominance of nodes in 
one state, meaning that a majority opinion appears via spontaneous symmetry breaking. 
While the symmetric active phase appears also in the CNVM with the rewire-to-same 
mechanism~\cite{bmin2018}, the asymmetric active phase happens exclusively with 
the rewire-to-random mechanism. It implies that the nonlinearity along with the
variation can produce a new phase and phase transition, with the slight modification 
in the details of the model.

\subsection{Coevolving Nonlinear Voter Model with Noise}

In social systems, noise is an inescapable factor. It arises from various aspects, such 
as the unpredictable nature of human interactions and the inherent randomness in individual 
choices. This stochasticity implies the possibilities to change 
the individual's state irrespective of their neighbors' states~\cite{diakonova2015}.
In this respect, the coevolutionary dynamics combining both nonlinearity and 
inherent noise was studied~\cite{raducha2020}. In this model, each node can 
change its state autonomously with a probability $\epsilon$, in addition to the 
dynamical rules of the CNVM. This probability produces a noise in coevolutionary dynamics.

In the CNVM with noise, there are three distinct phases similar to the original CNVM:
consensus, coexistence, and fragmentation.
However, noise prevents an absorbing or frozen state, turning the 
fragmentation and consensus states into dynamical states.
That is, the consensus and fragmentation states are no longer absorbing states 
but become dynamically steady states. The similar patterns are observed 
in the coevolving linear  voter model with noise~\cite{diakonova2015}.

In addition, the coexistence phases can be further divided into 
two distinct subclasses.
In the coexistence phase for $q<1$, there is a clear divide between a fully-mixing phase 
and a structured coexistence phase. The fully-mixing phase is the same phase 
as a coexistence phase observed in the ordinary CNVM
However, the structured coexistence phase shows significantly lower the density 
of active link due to the existence of two large communities that shows 
highly homogeneous opinions internally.

In the consensus phase for $q>1$, there are also two distinct subclasses:
a strong consensus where most nodes in the system remain in the same state 
and an alternating consensus where the majority opinion switches in time.
When the CNVM is integrated with noise, the resulting network structure and 
node states can be highly diverse. It implies that 
the noise can be a source of the diverse patterns in the structure
and opinion evolution of social networks.

\subsection{Coevolving Nonlinear Voter Models on Multilayer Networks}


Many real-world complex systems, from living organisms and human societies to transportation 
networks and critical infrastructures, function through multiple layers of interacting 
networks~\cite{kivela2014,lee2015}.
Additionally, the synergy between these layers is vital to understand
and control the function of networked systems~\cite{bmin2014,bmin2014b}.
Networks with multiple layers can also influence the opinion dynamics, leading to 
emergent phenomena that can better reflect real-world systems. 
Therefore, some studies have been presented that extends the coevolutionary dynamics 
of opinion dynamics from a single layer to multilayer 
networks~\cite{diakonova2014,bmin2019,klimek2016}.

Among these, there is a model of multilayer coevolution of the nonlinear voter 
model with synchronization of nodes' state between different layers~\cite{bmin2019}.
In this model, at each step a layer and a node in the chosen layer are selected at random.
Let us call it node $i$.
Then, the coevolutionary rule is the same as the CNVM on a single layer~\cite{bmin2018}.
In addition, if the copying process has occurred, a node in the other layer that is 
interconnected to node $i$ via an inter-link between layers also
changes its state to become the same state as node $i$.
This synchronization step ensures the same state for nodes connected 
across different layers. In this model, in addition to the network's plasticity 
$p$ and nonlinearity $q$ that already exist in the CNVM, another parameter 
$K$ has been introduced to represent the density of interlinks.

When the two layers have the same plasticity $p$, the fragmentation transition 
occurs with a larger value of $p$ compared to that on a single layer.
As the density of interlink $K$ increases, the location $p_c$ at which the 
fragment transition occurs becomes delayed.
It means that multiple layer structures delay and suppress the fragmentation of networks. 
In addition, an asymmetric fragmented phase for $q>1$ and an active shattered phase 
for $q < 1$ appear when two layers of networks have different values of plasticity, which does not 
exist in the CNVM in a single layer. 
The asymmetric fragmented phase represents the both layers undergo fragmentation
but the sizes of largest components for different layers are different each other.
And, the active shattered phase represents a state where the network becomes shattered 
into many isolated nodes, but the value of magnetization remains at zero.
Such non-trivial results demonstrate that the introduction of
multiple layers can give rise to new types of complex structures and dynamics, enriching 
our understanding of coevolutionary dynamics.

\section{Summary and Outlook}

In this paper, we have explored the coevolving nonlinear voter models from various 
perspectives as a representative of coevolutionary dynamics with group interactions.
The	``nonlinearity'' in the model represents an interaction where a node engages 
as a group with all of its neighbors, rather than pair-wise interactions. 
We have examined several variants of the CNVM incorporating the rewiring with 
the triadic closure, rewire-to-random mechanism, noise of flipping nodes' state, and 
multilayer structures  in coevolutionary dynamics. Integrating the group interactions
and the various factors, we have found a rich variety of phases and phase transitions
for both network structure and nodes' state.

In addition to the approach discussed in this 
paper, group interactions have been studied from various contexts in network science, 
such as complex contagions in social and biological systems~\cite{watts2002,kook2021,lee2023}, 
cooperative epidemics~\cite{chen2013,cai2015,bmin2020}, or higher-order representations 
of networked systems~\cite{battiston2021,battiston2020,majhi2022}.
Despite the advances in understanding group interactions, there is a still gap 
in exploring group interactions within the framework of coevolving dynamics. 
Therefore, it underlines an imperative need for further research in this area, given 
coevolutionary dynamics is one of the key factors in complex systems~\cite{complex}.
As a representative, the series of research related to the CNVM can provide a guide for 
further studies on coevolutionary dynamics considering diverse forms of group interactions.

Finally, we discuss a few recent advancements and future outlook on issues that are related 
to coevolutionary dynamics of group interactions. One active line of research related 
to group interaction is higher-order networks~\cite{battiston2021,battiston2020,majhi2022}.
Therefore, a straightforward extension of a coevolving model incorporating group 
interactions is adaptive dynamics on higher-order networks such as simplicial 
complexes~\cite{horstmeyer2020} and hypergraphs~\cite{papanikolaou2022, golovin2023}.
There are still more topics that need further research with the CNVM
including multi-opinion versions~\cite{shi2013}, dynamics in directed 
links~\cite{zschaler2012}, and aging effects~\cite{peralta2020}, to name a few.
From a broader perspective, the CNVM is a specific realization of coevolutionary 
dynamics with group interactions. The general effects of group interactions 
on coevolutionary dynamics, if any, still remain an open question. 
The CNVM offers a valuable framework for understanding coevolutionary dynamics, the 
research of group interactions combining coevolutionary dynamics presents vast 
opportunities for further exploration and refinement.

\section*{Acknowledgments}

We thank J. Choi and M. San Miguel for helpful discussion.
This work was supported by the National Research Foundation of Korea (NRF) 
grant funded by the Korea government (MSIT) (no.~2020R1I1A3068803).


\end{document}